\documentclass[12pt]{iopart}

\usepackage{iopams}  

\expandafter\let\csname equation*\endcsname\relax
\expandafter\let\csname endequation*\endcsname\relax

\usepackage{amsmath}        
\usepackage{graphicx} 
\usepackage{color}			
\usepackage{bm}			
\usepackage{cancel}			
\usepackage{empheq}

\DeclareMathOperator*{\argmax}{arg\,max}
\DeclareMathOperator*{\argmin}{arg\,min}

\begin{document}

\title[Hidden Markov models for  stochastic thermodynamics]{Hidden Markov models for  stochastic thermodynamics}

\author{John Bechhoefer}

\address{Department of Physics, Simon Fraser University, Burnaby, BC  V5A 1S6, Canada}
\ead{johnb@sfu.ca}
\vspace{10pt}
\begin{indented}
\item[]April 29, 2015
\end{indented}

\begin{abstract}
The formalism of state estimation and hidden Markov models (HMMs) can simplify and clarify the discussion of stochastic thermodynamics in the presence of feedback and measurement errors.  After reviewing the basic formalism, we use it to shed light on a recent discussion of phase transitions in the optimized response of an information engine, for which measurement noise serves as a control parameter.  The HMM formalism also shows that the value of additional information displays a maximum at intermediate signal-to-noise ratios.  Finally, we  discuss how systems open to information flow can apparently violate causality; the HMM formalism can quantify the performance gains due to such violations.
\end{abstract}

%
\vspace{2pc}

\submitto{\NJP}
%
%
%

\section{Introduction}
\label{sec:intro}

In 1867, at the dawn of statistical physics, Maxwell imagined a thought experiment that has both troubled and inspired physicists ever since \cite{leff03}.  In modern language, the issue is that  traditional thermodynamics posits a strict separation between observable macroscopic motion (dynamical systems) and unobservable degrees of freedom (heat).  But imagine---as can now be done experimentally on small systems where fluctuations are important---that it is possible to observe some of these hidden degrees of freedom.  (Maxwell's thought experiment used a ``demon" to accomplish the same task.)  In any case, the entropy of the system is reduced, and one can use the lower entropy to extract work from the surrounding heat bath, in seeming violation of the Second Law of thermodynamics.  

This blurring of macroscopic and microscopic degrees of freedom has led to a new field, \textit{stochastic thermodynamics}, which clarifies how thermodynamics should be applied to small systems where fluctuations are observable and  important \cite{seifert12}.  As we will see below, the nature of information acquired about the fluctuations---especially the precision with which they are measured and the time they become available---is of great importance.  Indeed, information is itself a thermodynamic resource, and  stochastic thermodynamics can be extended to accommodate the acquisition, dissipation, flow, and feedback of information \cite{touchette04,cao09,sagawa10,granger11,sagawa12,mandal12,deffner13,barato14b,barato14,hartich14,allahverdyan09b,horowitz13,horowitz14}.  For a recent review, see \cite{parrondo15}.  

The goal of the present contribution is to combine ideas from control theory (state estimation) \cite{bechhoefer05,astrom08} with ideas from computer science about hidden Markov models \cite{press07,durbin98,rabiner89,murphy12,cappe05} in order to explain some recent surprising observations from stochastic thermodynamics about how Maxwell's demon operates in the presence of measurement errors \cite{bauer14}.  As a bonus, the formalism we discuss suggests a number of interesting areas where the stochastic thermodynamics of information may be extended.

\section{Coarse graining and discrete state spaces}
\label{sec:hmm-coarse-graining}

\begin{figure}[ht]
\begin{center}
	\includegraphics[width=6.0in]{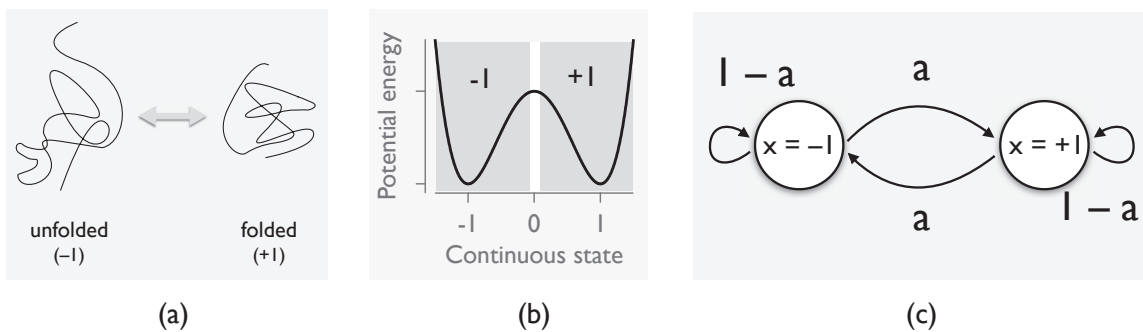}	
	\caption{Coarse graining to find a Markov model.  (a)  A protein in water alternates between two conformations.  (b)  A one-dimensional projection of the dynamics.  White vertical line denotes threshold separating the $\pm 1$ states.  (c)  Graphical depiction of a symmetric two-state Markov chain.}
	\label{fig:markov}
\end{center}
\end{figure}

In the simplest non-trivial example of a discrete state space, a state $x$ can, at each discrete time point, take on one of two values, for example $-1$ and $+1$.  While systems such as spin-$\tfrac{1}{2}$ particles are inherently discrete, a broad range of physical systems---even classical, continuous state spaces---can often be well approximated by discrete systems after coarse graining.  Figure~\ref{fig:markov}(a) sketches such a system, a protein in solution that alternates  between a loose unfolded ($-1$) and a compact folded ($+1$) state.  Other biological examples of two-state systems include ion channels that can be open or closed, gene-transcription repressor sites that can be occupied or empty, and sensory receptors that can be active or silent (chapter 7 in \cite{phillips12}).

Figure~\ref{fig:markov} illustrates schematically how to coarse grain from a physical situation, such as a protein in water, to a discrete-time Markov model.  In (a), we depict two states of the protein, labeled ``unfolded" and ``folded" or, equivalently, $-1$ and $+1$.  The word ``state" is here a shorthand for ``macrostate" and is associated with many microstates, each of which corresponds to a slightly different protein conformation that preserves the general property in question.  In (b), we project the full dynamics onto a one-dimensional subspace modeled by a double-well potential.  States with $x<0$ are classified as $-1$, and states with $x>0$ are classified as $+1$.  The symmetry of the potential implies that the protein spends equal time in the two states, which is a special situation.  In (c), we show a graphical depiction of the discrete, two-state Markov chain dynamics, where in a time $\tau$, states remain the same with probability $1-a$ and hop to the other with probability $a$.  In order for a two-state description to reasonably approximate the dynamics, the dwell time spent in each well must be much longer than the time scale for fast motion within a well.  This holds when a single energy barrier $E_{\rm b}$ separates two states and whose  height is much larger than $kT$.

Why might we want to approximate physical systems by discrete state spaces?  
\begin{itemize}
\item \textit{Clarity}:  We can isolate just the important degrees of freedom, letting the others be uncontrolled and even unobserved.  

\item \textit{Simplicity}:  The mathematical description is more straightforward. 

\item \textit{Generality}:  Any dynamics that can be modeled on a computer is necessarily discretized in both time and state.

\end{itemize}

\section{Markov chains}
\label{sec:hmm-markov}

\begin{figure}[ht]
\begin{center}
	\includegraphics[width=5.0in]{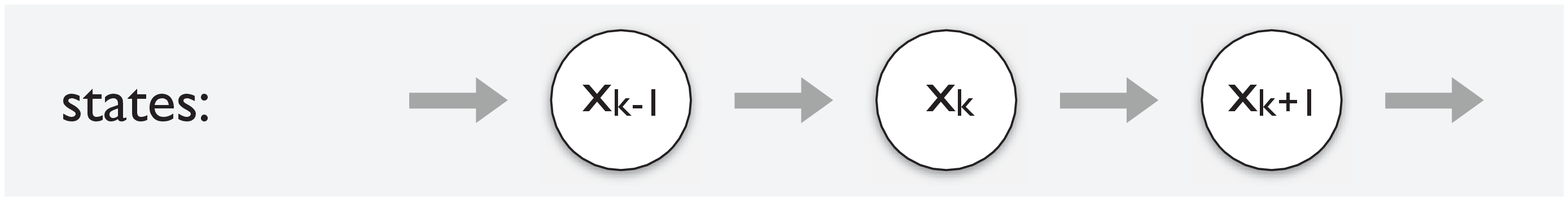}	
	\caption{Markov model graphical structure. The state $x_{k+1}$ depends only on $x_k$.}
	\label{fig:graphicalMarkov}
\end{center}
\end{figure}

Let us briefly recall the basics of  discrete-state-space systems in discrete time.   Consider a system described at time $k$ by a state $x_k$ that can be in one of $n$ possible states, indexed by the values $1$ to $n$.  The index is distinguished from its \textit{value}, which, for a two-state system, might be $\{\pm 1 \}$, $\{0,1\}$, or even $\{$left, right$\}$.  Let $P(x_k = i)$ be the probability that, at time $k$, the system is in the state indexed by $i$.  The distribution is normalized by enforcing $\sum_{i=1}^n P(x_k = i) = 1$ or, more succinctly, $\sum_{x_k} P(x_k)=1$.  For dynamics, we consider \textit{Markov chains}, which are systems with discrete time and discrete  states.  The Markov property implies that the next state depends only on the current state, as illustrated graphically in figure~\ref{fig:graphicalMarkov}, which may be compared with figure~\ref{fig:markov}(c).

For Markov chains, the dynamics are specified in terms of an $n \times n$ \textit{transition matrix} $\mathbf{A}$ whose elements $A_{ij} \equiv P(x_{k+1} = i \, | \, x_k = j)$ satisfy $0 \le A_{ij} \le 1$.  That is, $A_{ij}$ gives the rate of $j \to i$ transitions.  For example, a general two-state system has 
\begin{equation}
	\mathbf{A} = \begin{pmatrix} 1-a_0 & a_1 \\ a_0 & 1-a_1 \end{pmatrix} \,.
\label{eq:hmm-markov-transition-matrix}
\end{equation}
Notice that the columns of $\mathbf{A}$ sum to 1, as required by the normalization of probability distributions.  In words, if you start in state $j$ then you must end up in one of the $n$ possible states, indexed by $i$.  Figure~\ref{fig:markov}(c) depicts \eref{eq:hmm-markov-transition-matrix} graphically, with $a_0=a_1=a$.  A matrix with elements $0 \le A_{ij} \le 1$ and $\sum_i A_{ij}=1$ is a (left) \textit{stochastic matrix}.

Define the $n$-dimensional \textit{stochastic vector} $\mathbf{p}_k$, whose elements $p^{(j)}_k \equiv P(x_k = j)$ give the probability to be in state $j$ at time $k$.  Then $0 \le p^{(j)}_k \le 1$ and $\sum_j p^{(j)}_k  = 1$ and 
\begin{equation}
	p^{(i)}_{k+1} = \sum_{j=1}^n P(x_{k+1}=i,x_k=j) 
		= \sum_{j=1}^n \underbrace{P(x_{k+1}=i \, | \, x_k=j)}_{A_{ij}} \, P(x_k=j) 
		= \sum_{j=1}^n A_{ij} \, p^{(j)}_k \,.
\label{eq:hmm-discrete-master}
\end{equation}
More compactly, $\mathbf{p}_{k+1} = \mathbf{A} \, \mathbf{p}_k$, a linear difference equation with solution $\mathbf{p}_k = \mathbf{A}^k \,  \mathbf{p}_0$ known as the \textit{discrete-time master equation}.  Often, we seek the steady-state distribution, defined by $\mathbf{p} = \mathbf{A} \mathbf{p}$.  One way to find $\mathbf{p}$ is to repeatedly iterate \eref{eq:hmm-discrete-master}; another is to note that the steady-state distribution of probabilities corresponds to the eigenvector associated with an eigenvalue equal to 1.  A stochastic matrix must  have such an eigenvalue, since $\mathbf{A}-\mathbf{I}$ is a matrix whose columns all sum to zero.  They are then linearly dependent, with zero determinant.

%

For example, the two-state Markov model with transition matrix $\mathbf{A}$ given by \eref{eq:hmm-markov-transition-matrix} has eigenvalues $\lambda = 1$ and $1-(a_0+a_1)$.  The normalized eigenvector corresponding to $\lambda=1$ is  
\begin{equation}
	\mathbf{p}^* = \tfrac{1}{a_0+a_1} \, \begin{pmatrix} a_1 \\ a_0 \end{pmatrix} \,.
\end{equation}
For the symmetric case, $a_0=a_1 \equiv a$ and $\mathbf{p}^* = \bigl( \begin{smallmatrix} 0.5 \\[2pt] 0.5 \end{smallmatrix} \bigr)$, independent of $a$.  By symmetry, both states are \textit{a priori} equally probable.

\section{Hidden Markov models}
\label{sec:hmm-hidden-markov}

\begin{figure}[ht]
\begin{center}
	\includegraphics[width=6.0in]{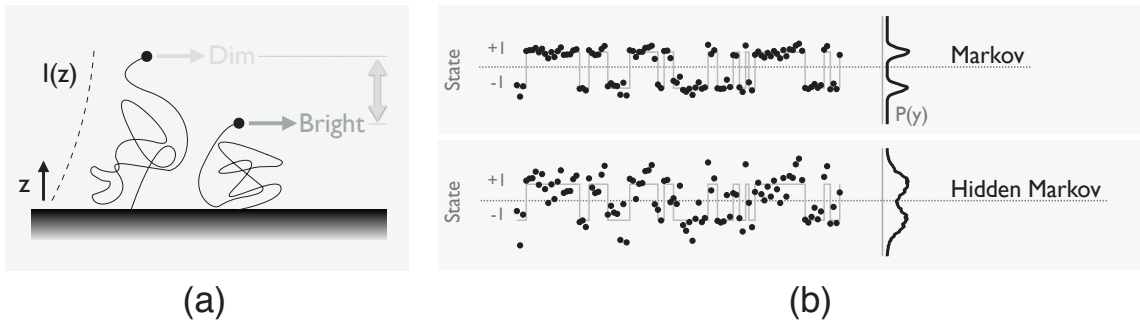}	
	\caption{Markov vs.~hidden Markov models.  (a)  Schematic illustration of a scattering probe of protein conformation where evanescent-wave illumination changes the intensity of scattered light in the two states.  (b)  Observations for a two-state Markov process where observations correlate  unambiguously with states (top) and a hidden Markov process (bottom) where conditional distributions overlap.  True state in light gray.  Observations $y_k$ are indicated by round markers and have Gaussian noise, with standard deviation $\sigma=0.2$ (top) and $0.6$ (bottom).  Histograms of $y_k$ are compiled from $10^4$ observations, with 100 shown.}
	\label{fig:hmmFig}
\end{center}
\end{figure}

Often, the states of a Markov chain are not directly observable; however, there may be measurements (or \textit{emitted symbols}) that correlate with the underlying states.  The combination is known as a \textit{hidden Markov model} (HMM).  The hidden states are also sometimes known as \textit{latent} variables \cite{murphy12}.  The observations are assumed to have no memory:  what is measured depends only on the current state, and nothing else.  The graphical structure of an HMM is illustrated in figure~\ref{fig:hmmFig}.

\begin{figure}[ht]
\begin{center}
	\includegraphics[width=5.0in]{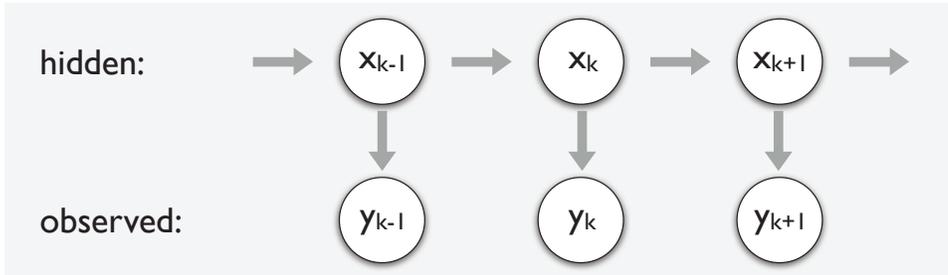}	
	\caption{HMM graphical structure. The states $x_k$ form a Markov process that is not directly observable. The observations $y_k$ depend only on $x_k$.}
	\label{fig:graphicalHMM}
\end{center}
\end{figure}

In the example of proteins that alternate between unfolded and folded states, the molecule itself is  not directly observable.  One way to observe the configuration is to attach a particle to one end of the protein and anchor the other end to a surface \cite{zocchi96}, as illustrated in figure~\ref{fig:hmmFig}(a).  As the protein folds and unfolds, the particle moves up and down from the surface.  We can illuminate the region near the surface using an evanescent wave via the technique known as total internal reflection microscopy.  The intensity $I(z)$ of light scattered by the bead at height $z$ from the surface will decrease exponentially as $I(z) \propto e^{-z/z_0}$, with $z_0 \approx 100$ nm.  The two states will then correspond to two different scattering intensities.  The observation $y_k$ is the number of recorded photons, integrated over a time that is shorter than the dwell time in each local potential well. 

As with states, we can further simplify by discretizing the intensities, classifying as ``dim" intensities below a given threshold and ``bright" intensities above that threshold.  ``Dim" and ``bright" then become two observation \textit{symbols}.  Because light scattering is itself a stochastic process, the protein can be in one state but emit the ``wrong" symbol, as illustrated in figure~\ref{fig:hmmFig}(b).  We can describe such a situation by defining the observations $y_k=\pm 1$ and noting that they are related to the states probabilistically via an \textit{observation matrix} $\mathbf{B}$ having components $B_{ij} \equiv P(y_k = i | x_k = j)$:
\begin{equation}
	\mathbf{B} =  \begin{pmatrix} 1-b & b \\ b & 1-b \end{pmatrix} \,,
\label{eq:hmm-observation-matrix}
\end{equation}
where we suppose, for simplicity, that errors are symmetric.  Because observations have no memory, the probability to observe $y_k$ depends only on the current state $x_k$. 
 
In words, the matrix $\mathbf{B}$ states that an observation is correct with probability $1-b$ and wrong with probability $b$.  Like the transition matrix $\mathbf{A}$, the matrix $\mathbf{B}$ is stochastic, with columns that sum to 1.  Its rows also sum to 1, but only because of the symmetry between states.   Note that the number of observation symbols, $m$, need not equal the number of internal states, $n$.  The $m \times n$ matrix $\mathbf{B}$ can have $m$ bigger or smaller than $n$.  The case of continuous observations ($m \to \infty$) is also straightforward.  Larger values of $m$  increase knowledge of the underlying state somewhat.  

One interesting feature of HMMs is that states $x_k$ follow a Markov process and so does  the combined process for $x_k$ and $y_k$, but not necessarily the observations $y_k$.  The analysis of HMMs is thus more difficult than for ordinary Markov processes.

The literature on HMMs is both vast and dispersed.  For treatments of increasing complexity, see section 16.3 of \textit{Numerical Recipes} \cite{press07}, the bioinformatics book by Durbin et al.~\cite{durbin98}, a classic tutorial from the speech-recognition literature \cite{rabiner89}, the control-influenced book by S\"arkk\"a \cite{sarkka13}, and the mathematical treatment of Capp\'e et al.~\cite{cappe05}.  The tutorial by Rabiner  \cite{rabiner89} has been particularly influential; however, its notation and ways of deriving results are more complicated than need be, and some of its methods have been replaced by better algorithms.  The discussion here is based largely on the cleaner derivations in \cite{sarkka13}.

\section{State estimation}
\label{sec:hmm-state-estimation}

Hidden Markov models are specified by a transition matrix $\mathbf{A}$ and observation matrix $\mathbf{B}$.  Let us pose the following problem:  Given the output of a hidden Markov model (HMM), what can be inferred about the states?  The answer depends both on the information available and the exact quantity desired.  Here, we focus on two cases:

\begin{enumerate}
\item \textit{Filtering}, or $P(x_k | y^k)$.  We estimate the probabilities for each state based on observations $y^k \equiv \{y_1, y_2, \ldots, y_k \}$ up to and including the present time $k$.  Filtering is appropriate for real-time applications such as control.\footnote{
An alternate notation for $P(x_k | y^k)$ is $P(x_k | y_{1:k})$.  Our notation seems cleaner and easier to read.}

\item \textit{Smoothing}, or $P(x_k | y^N)$, for $N>k$.  Smoothing uses data from the future as well as the past in the offline post-processing of $N$ observations.  

\end{enumerate}

Another quantity of interest is the \textit{most likely path}, defined as $\argmax_{x^N} \, P(x^N | y^N)$, which may be found by an algorithm due to Viterbi \cite{press07}.  For example, McKinney et al.~study transitions between different configurations of a DNA Holliday junction, using fluorescence resonance energy transfer (FRET) to read out the states, and infer the most likely state sequence \cite{mckinney06}.  Since path estimates are less useful for feedback control, we will consider them only in passing, in section~\ref{sec:hmm-phase-trans}.  We will also see that smoothing estimates provide a useful contrast with filter estimates.

\subsection{Filtering}
\label{sec:hmm-filtering}

The filtering problem is to find the probability distribution of the state $x_k$ based on the past and current observations $y^k$ from time 1 to time $k$.  We assume that the dynamics have been coarse grained to be Markov, so that the state $x_{k+1}$ depends only on the state $x_k$.  Then $P(x_{k+1} | x^k,\cancel{y^k}) = P(x_{k+1} | x_k)$, where the ``cancel" slash indicates conditional independence:  conditioning on $x_k$ ``blocks" the influence of all other variables.  The $x^{k-1}$ are blocked, too:  the state at time $k+1$ depends \textit{only} on the state at time $k$.

From marginalization and the definition of conditional probability, we have
\begin{align}
	P(x_{k+1} | y^k) &= \sum_{x_k} \, P(x_{k+1}, x_k | y^k) \nonumber \\
	&= \sum_{x_k} \, P(x_{k+1} | x_k, \cancel{y^k}) \, P(x_k | y^k) \nonumber \\
	&= \sum_{x_k} \, P(x_{k+1} | x_k) \, P(x_k | y^k) \,.
\label{eq:hmm-Bayesian-filter-predict}
\end{align}
Equation~\eref{eq:hmm-Bayesian-filter-predict} predicts the state $x_{k+1}$ on the basis of $y^k$, assuming that the previous filter estimate, $P(x_k | y^k)$ is already known.  Once the new observation $y_{k+1}$ is available, we can use Bayes' Theorem and the memoryless property of observations, $P(y_k | x^k, \cancel{y^{k-1}}) = P(y_k | x_k)$, to update the prediction \eref{eq:hmm-Bayesian-filter-predict} to incorporate the new observation.  Then,
\begin{equation}
	P(x_{k+1} | y^{k+1}) 
		= \frac{1}{Z_{k+1}} \, P(y_{k+1} | x_{k+1}, \cancel{y^k}) \, P(x_{k+1} | y^k) \,,
\label{eq:hmm-Bayesian-filter-update}
\end{equation}
where $Z_{k+1}$ normalizes the distribution.  Equations \eref{eq:hmm-Bayesian-filter-predict}--\eref{eq:hmm-Bayesian-filter-update} constitute the Bayesian filtering equations \cite{sarkka13,ho64}.  Because of their importance, we collect them here:
\begin{empheq}[box=\fbox]{align}
	P(x_{k+1} | y^k) &= \sum_{x_k} \, P(x_{k+1} | x_k) \, P(x_k | y^k) &&\text{predict} 
		\nonumber \\[3pt]
	&\, \downarrow &&\hspace{9pt}\downarrow \nonumber \\[3pt]
	P(x_{k+1} | y^{k+1}) &= \frac{1}{Z_{k+1}} \, P(y_{k+1} | x_{k+1}) \, P(x_{k+1} | y^k) 
		&& \text{update} \,.
\label{eq:hmm-Bayesian-filter}
\end{empheq}
The normalization (\textit{partition function}) $Z_{k+1}$ is given by
\begin{align}
	Z_{k+1} = P(y_{k+1} | y^k) &= \sum_{x_{k+1}} P(y_{k+1} | x_{k+1}) \, P(x_{k+1} | y^k) \,.
\label{eq:hmm-partition-function}
\end{align}

Note that the HMM literature, e.g., \cite{press07} and \cite{durbin98} , expresses 
\eref{eq:hmm-Bayesian-filter} differently, using joint probabilities such as $P(x_k,y^k)$ rather than conditional probabilities such as $P(x_k | y^k)$.  Using joint probabilities leads to the \textit{forward algorithm}.  Our notation emphasizes the similarities between HMM and state-space models of dynamics; the formulas of one apply mostly to the other, with  $\sum_{x_k} \, \leftrightarrow  \int dx_k$.  For continuous state spaces with linear dynamics and Gaussian noise, \eref{eq:hmm-Bayesian-filter} is equivalent to the \textit{Kalman filter} \cite{sarkka13}.  Below, we will see that using conditional probabilities also has numerical advantages.

\begin{figure}[ht]
\begin{center}
	\includegraphics[width=6.0in]{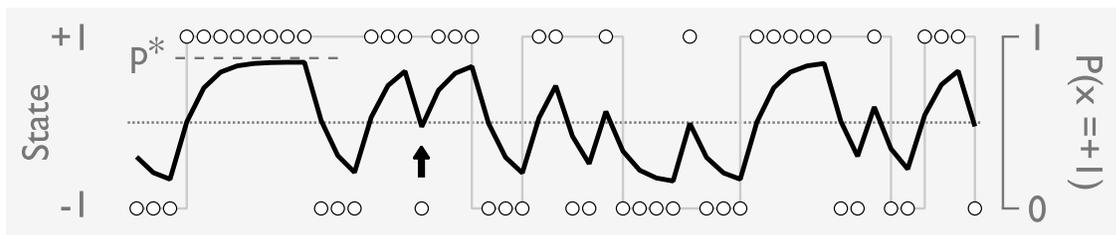}	
	\caption{Filtering for a symmetric, two-state, two-symbol hidden Markov model with $a=0.2$ and $b=0.3$.  Light gray line shows true state, which is hidden.  Markers show 100 observations.  Heavy black line shows the probability that the state equals $+1$, given by $P(x_k=1|y^k)$.  The maximum confidence level $p^* \approx 0.85$ (dashed line).}
	\label{fig:hmm-hmmInfFilter}
\end{center}
\end{figure}

Figure~\ref{fig:hmm-hmmInfFilter} shows filtering in action for a symmetric, two-state, two-symbol hidden Markov model.  The time series of observations $y_k$ (markers) disagrees with the true state 30\% of the time.  The black line shows $P(x_k = 1 | y^k)$.  When that probability is below the dashed line at $0.5$, the most likely state is 0.  For the value of $a$ used in the dynamic matrix ($a=0.2$), the filter estimate $x^{(f)}_k =$ arg max$_{x_k}$ $P(x_k | y^k)$ disagrees with the observation only 7\% of the time, a noticeable improvement over the naive 30\%.  Notice that whenever the state changes, the filter probability responds, with a time constant set by both observational noise ($b$) and dynamics ($a$).  A long string of identical observations causes filter confidence to saturate at $p^*$ (dashed line).

There is an advantage to recording the probability estimates (black line) rather than simply the MAP (maximum a posteriori) estimate, which here is just the more likely of the two possibilities.  When the filter is wrong, the two probabilities are often not that different.  An example is indicated by the arrow in figure~\ref{fig:hmm-hmmInfFilter}.  Thus, marginalizing (averaging) any prediction over all possibilities rather than just the most likely will improve estimates.  Of course, a string of wrong symbols can fool the filter.  See, in figure~\ref{fig:hmm-hmmInfFilter}, the three wrong symbols just to the left of the arrow.  

Below, we will see that the filtered estimate becomes significantly more reliable as $a \to 0$.  Intuitively, small $a$ means that states have a long \textit{dwell time}, so that averaging observations over times of the order of the dwell time can reduce the effect of the observational noise, which is quantified by the parameter $b$.

\subsection{Smoothing}
\label{sec:hmm-smoothing}

If we estimate the state $x_k$ after gathering $N$ observations ($N>k$), we can use the ``future" information to improve upon the filter estimate.  In the control-theory literature, such estimates are called ``smoother" estimates, as they further reduce the consequences of observation noise.

The smoother estimate has two stages.  First, we use the filter algorithm \eref{eq:hmm-Bayesian-filter} to calculate $P(x_k|y^k)$ and $P(x_{k+1} | y^k)$ for each $k \in [1,N]$.  Then we calculate $P(x_k | y^N)$ via a backward recursion relation from the final time $N$ to the initial time $1$.
\begin{empheq}[box=\fbox]{align}
	P(x_k | y^N) = P(x_k | y^k) \, \sum_{x_{k+1}} \, \frac{P(x_{k+1} | x_k) \, 
		P(x_{k+1} | y^N)} {P(x_{k+1} | y^k)} \,.
\label{eq:hmm-backward-alg}
\end{empheq}
The backwards recursion relation is initialized by $P(x_N,y^N)$, the last step of the forward filter recursion.

To derive \eref{eq:hmm-backward-alg}, we introduce the state $x_{k+1}$, which we will remove later by marginalization \cite{sarkka13}.  Thus, 
\begin{align}
	P(x_k , x_{k+1} | y^N) = P(x_k | x_{k+1}, y^N) \, P(x_{k+1} | y^N) \,.  
\label{eq:smoother-derivation0}
\end{align}
But
\begin{align}
	P(x_k | x_{k+1}, y^N) 	
	= P(x_k | x_{k+1}, y^k) 
	= \frac{P(x_k, x_{k+1} | y^k)} 
		{P(x_{k+1} | y^k)} 
	=  \frac{P(x_{k+1} | x_k, \cancel{y^k}) \, 
		P(x_k | y^k)} 
		{P(x_{k+1} | y^k)}  \,,
\label{eq:smoother-derivation}
\end{align}
using conditional probability and the Markov property.  Substituting into \eref{eq:smoother-derivation0},  
\begin{align}
	P(x_k , x_{k+1} | y^N) 
	= \frac{P(x_k | y^k) \, P(x_{k+1} | x_k) \, 
		 P(x_{k+1} | y^N)} {P(x_{k+1} | y^k)} \,.
\label{eq:smoother-not-marginalized}
\end{align}
Summing both sides over $x_{k+1}$ gives \eref{eq:hmm-backward-alg}.  

The algorithm defined by \eref{eq:hmm-Bayesian-filter} and \eref{eq:hmm-backward-alg} is equivalent to the \textit{Rauch-Tung-Striebel smoother} from control theory when applied to continuous state spaces, linear dynamics, and white-noise inputs~\cite{sarkka13}.  In the HMM literature, a close variant is the \textit{forward-backward algorithm}~\cite{durbin98}.

\begin{figure}[ht]
\begin{center}
	\includegraphics[width=6.0in]{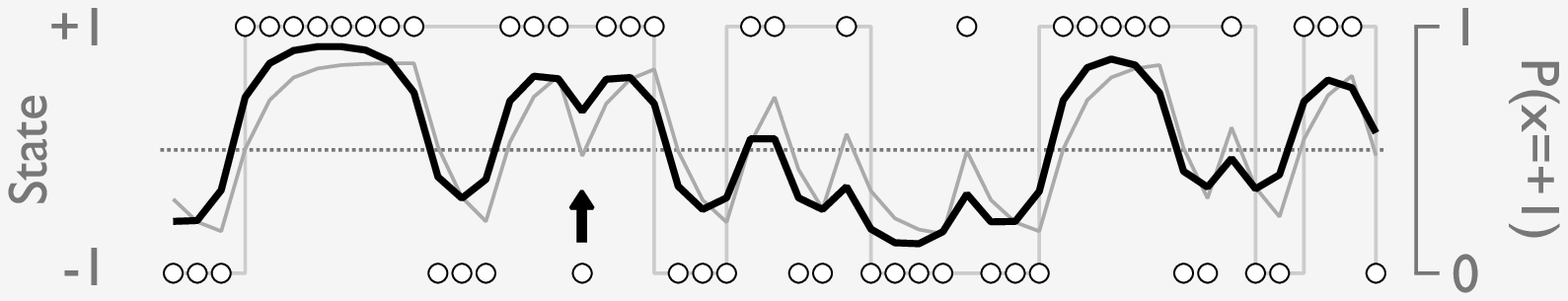}	
	\caption{Smoother estimates (black line) for two-state, two-symbol HMM with $a=0.2$ and $b=0.3$.  Filter estimate is shown as a light gray trace.  The simulation and filter estimate are both from figure~\ref{fig:hmm-hmmInfFilter}.}
	\label{fig:hmm-hmmInfSmoother}
\end{center}
\end{figure}

We can apply the smoother algorithm to the example of section~\ref{sec:hmm-filtering} and obtain similar results.  In figure~\ref{fig:hmm-hmmInfSmoother}, we plot the smoother estimate, with the filter estimate added as a light gray trace.  Despite their similarity, the differences are instructive:  The filter always lag (reacts) to observations, whereas the smoother curve is more symmetric in time.  Flipping the direction of time alters the overall form of the filter plot but not the smoother.  
The smoother estimates are more confident than the filter estimates, as they use more information.  Look at the time step indicated by the arrow.  The filter estimate is just barely mistaken, but the smoother estimate makes the correct call, aided by the three correct observations that come before and the three after.

The phase lag apparent in the filter estimate is consistent with causality.  Indeed, for continuous state spaces, the well-known Bode gain-phase relations---the ``magnitude-phase" equivalent of the Kramers-Kronig relations \cite{bechhoefer11}---give the minimum phase lag for the output of a dynamical system that is consistent with causality.  The smoother estimate in figure~\ref{fig:hmm-hmmInfSmoother} has \textit{zero} phase lag, as expected since it uses past and future information equally.  Sudden jumps are anticipated by the smoother \textit{before}  they happen. 

Intuitively, an estimator that uses more information should perform better.  We can formalize this intuition via the notion of conditional \textit{Shannon entropy} \cite{cover06}.  With $p_j \equiv P(x_k = j \, | \, y^k)$, 
\begin{equation}
	H(x_k | y^k) \equiv -\sum_{j=1}^n p_j \, \log p_j \,,
\label{eq:shannon-entropy-def}
\end{equation}
where using a base-2 logarithm gives units of \textit{bits}.  For large-enough $k$, the average of $H(x_k | y^k)$ over $y^k$ becomes independent of $k$. Averaging over a single long time series of observations then leads to $\langle H(x_k | y^k) \rangle = H(x|\overleftarrow{y})$, where $\overleftarrow{y}$ denotes past and present observations.   A similar definition holds for the smoother entropy, $H(x_k | y^N)$ and leads to a steady-state smoother entropy $H(x|\overleftrightarrow{y})$, where $\overleftrightarrow{y}$ includes both past and future observations.  To characterize the performance of filtering and smoothing, we recall that for a two-state probability distribution, the entropy ranges from 1 bit (equal probabilities for each possibility) to 0 bits (certainty about each possibility).

\begin{figure}[ht]
\begin{center}
	\includegraphics[width=5.0in]{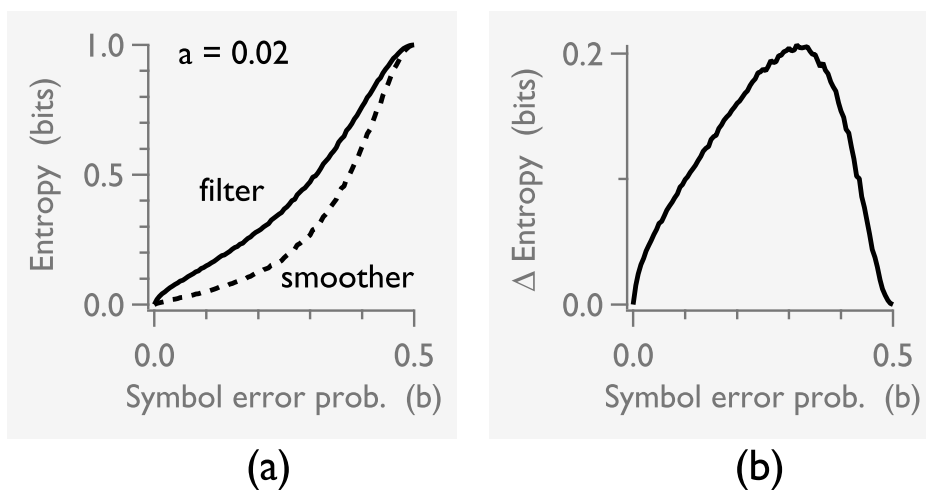}	
	\caption{Smoother outperforms filter.  (a) Shannon entropies of filter and smoother state estimates.  The symmetric transition matrix $\mathbf{A}$ has parameter $a=0.02$.  (b)  Filter minus smoother.  Calculations use time series of length $10^5$.}
	\label{fig:FiltervSmoother}
\end{center}
\end{figure}

Figure~\ref{fig:FiltervSmoother}(a) shows the steady-state filter and smoother Shannon entropies as a function of $b$, the error rate in the observation matrix $\mathbf{B}$.  At small values of $a$, the smoother has a greater advantage relative to the filter: when dwell times in each state are long, the information provided by averaging is more important.  Figure~\ref{fig:FiltervSmoother}(b) plots the difference between filter and smoother entropies.  For $b=0$, the difference vanishes: with no noise, the observation perfectly determines the state, and there is no uncertainty about it afterwards.  For $b=0.5$, the observations convey no information, and $H(x|\overleftarrow{y}) = H(x|\overleftrightarrow{y}) = H(x) = 1$ bit and the difference is again zero.  For intermediate values of $b$, the smoother entropy is lower than the filter entropy.

\section{Learning hidden Markov models}
\label{sec:hmm-baum-welch}

The state-estimation procedures described above assume that the transition matrix $\mathbf{A}$, the emission matrix $\mathbf{B}$, and initial probability $P(x_1)$ are known.  If not, they can be estimated from the observations $y^N$.   In the context of HMMs, the task is called, variously, \textit{parameter inference}, \textit{learning}, and \textit{training} \cite{press07}.  In the control-theory literature on continuous state spaces, it is known as \textit{system identification} \cite{ljung99}.

The general approach is to maximize the likelihood of the unknown quantities, grouped here into a single parameter vector $\bm{\theta}$.  That is, we seek
\begin{equation}
	\bm{\theta}^* = \argmax_{\bm{\theta}} P(y^N | \bm{\theta}) 
		= \argmin_{\bm{\theta}} \left[ - \ln P(y^N | \bm{\theta}) \right] \,,
\label{eq:hmm-max-likelihood-pars}
\end{equation}
where it is better to compute $L(\bm \theta) \equiv -\ln P(y^N | \bm{\theta})$ because $P(y^N | \bm{\theta})$ decreases exponentially with $N$, leading to numerical underflow.  The negative sign is a convention from  least-squares curve fitting, where $\chi^2(\bm{\theta})$ is also proportional to the negative log likelihood of the data \cite{press07}.  

We can find the total likelihood $P(y^N | \bm{\theta})$ from the normalization condition in  \eref{eq:hmm-Bayesian-filter}:
\begin{equation}
	P \left(y^N \right) = \underbrace{\prod_{k=1}^N  P \left( y_k | y^{k-1} \right)}_{\rm chain \, rule} = \prod_{k=1}^N Z_k \,,
\label{eq:hmm-total-likelihood}
\end{equation}
where $Z_1 \equiv P(y_1)$.  Then
\begin{equation}
	L(\bm \theta) = -\sum_{k=1}^N \ln \sum_{x_k} P \left( y_k | x_k \right) \, 
		P \left( x_k | y^{k-1} \right) \,,
\end{equation}
where all right-hand-side terms depend also on $\bm{\theta}$.  Since $L (\bm{\theta})$ is just a function of $\bm{\theta}$, we can use standard optimization routines to find the $\bm{\theta}^*$ that minimizes $L$.  

In the HMM literature, an alternate approach to finding $\bm{\theta}^*$ is based on the Expectation Maximization (EM), or Baum-Welch algorithm \cite{murphy12,durbin98}.  In a two-step iteration, one finds $\bm \theta$ by maximum likelihood assuming that the hidden states $x^N$ are known and then infers states $x^N$ from the smoother algorithm assuming $\bm{\theta}$ is known. The algorithm converges locally but can be very slow.  Indeed, the EM algorithm can seldom compete against the more sophisticated direct-optimization algorithms readily available in standard scientific programming languages.  EM algorithms can, however, be the starting point for recursive variants that allow for adaptation \cite{krishnamurthy93}.  A third approach to finding HMM parameters, based on finding the most likely (Viterbi) path, can also converge faster than EM and be more robust \cite{allahverdyan11}.

\section{Control of discrete-state-space systems}
\label{sec:hmm-control}

\begin{figure}[ht]
\begin{center}
	\includegraphics[width=5.0in]{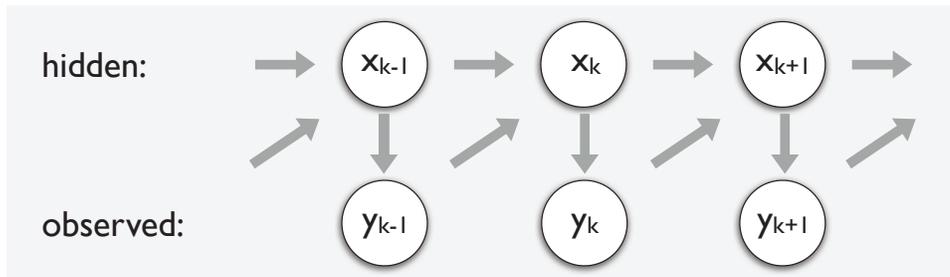}	
	\caption{Partially observable Markov decision process graphical structure. The hidden states $x_{k+1}$ form a Markov process whose transitions depend both on states $x_k$ and observations $y_k$.}
	\label{fig:graphicalPOMDP}
\end{center}
\end{figure}

We can now discuss the control of Markov models and HMMs.  In the context of discrete state spaces, the control $u_k$ influences the transition probability, which becomes $P(x_{k+1} | x_k, u_k)$ and is described by a time-dependent transition matrix $\mathbf{A}_k$ and a graphical structure illustrated in figure~\ref{fig:graphicalPOMDP}.  Note that our previous discussion of state estimation (filtering) never assumed that the transition matrix is time independent.

The control of Markov chains is formally known as a \textit{Markov Decision Process} (MDP), while that of HMMs is known as a \textit{Partially Observable Markov Decision Process} (POMDP).  Optimal-control protocols that minimize some cost function can be found using Bellman's dynamic programming, which is a general algorithm for problems involving sequential decisions \cite{press07,bertsekas05}.  In this setting, control is viewed as a blend of state estimation and decision theory \cite{bertsekas05,jensen07}.  The goal is to choose actions based on available information in order to minimize a cost function.

Here, we will present such ideas more informally, using a well-studied example:  optimal work extraction from a two-state system with noisy observations and feedback.  This problem is closely related to a famous thought experiment (recently realized experimentally \cite{toyabe10}), Maxwell's demon.

\subsection{Maxwell's demon}
\label{sec:hmm-maxwell-demon}

As discussed in the introduction, a Maxwell demon is a device where information about the state of a system is used to extract energy from a heat bath, in violation of the traditional form of the Second Law of thermodynamics.  How is this possible?  The catch is that we have assumed that information carries no cost.  A first attempt at resolving the paradox hypothesized that energy is dissipated in acquiring information  \cite{leff03}.  However, that turns out not to be true in general:  one can sometimes acquire information without doing work.  In its Kelvin-Planck formulation, the Second Law requires that no cyclic protocol of parameter variation can extract work from the heat bath of an equilibrium system held at constant temperature.  Specifying a cyclic protocol can be subtle.  Naively, a cyclic protocol requires that any potentials that are changed must be returned to their initial state; any mechanical part (pistons, etc.) that are moved must be moved back; and so on.  But it also applies to information.  In particular, any information acquired must be erased.  In 1961, Landauer proposed that the erasure step necessarily required energy dissipation of at least $kT \, \ln 2$ per bit, an amount that equals or exceeds the amount of work that can be extracted, thus saving (or extending) the Second Law \cite{landauer61}.  Landauer's prediction has recently been confirmed experimentally \cite{berut12,jun14}, as has its converse, the Szil\'ard engine, which uses acquired information to extract work from a heat bath \cite{toyabe10,koski14,roldan14}.  

\begin{figure}[ht]
\begin{center}
	\includegraphics[width=6.0in]{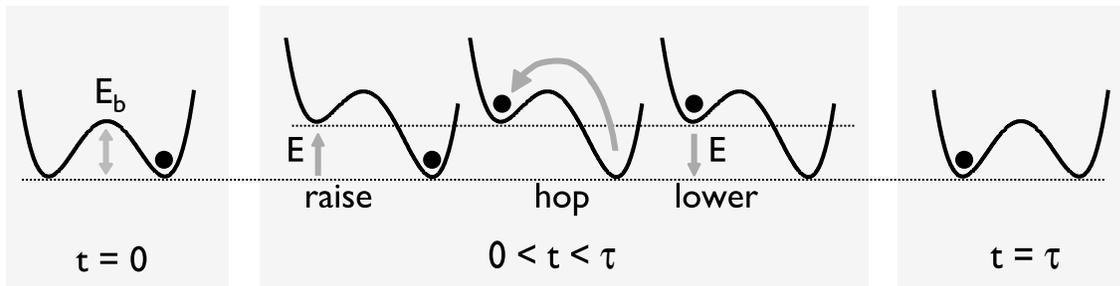}	
	\caption{Converting information to work in a two-state system that hops back and forth between ``left-well" and ``right-well" states separated by a high energy barrier $E_{\rm b}$.  If the system is observed to be in its right-well state, then we can raise the left well without doing work.  After a time $\tau$, the well is lowered.  If the left state is occupied, we extract an energy $E$ that can be used to perform work.}
	\label{fig:hmm-optimalWorkExtraction}
\end{center}
\end{figure}

\subsection{A simple model, with fully observed states}

We consider a particle in a fluid, subject to a double-well potential that may be manipulated by the experimenter (figure~\ref{fig:hmm-optimalWorkExtraction}).  It is a useful setting for thinking about the issues raised by a Maxwell demon and is a situation that can now be realized experimentally \cite{berut12,jun14}.  We assume that the energy barrier is large ($E_{\rm b} \gg kT$), so that we can coarse grain to two-state Markov dynamics, as discussed in section~\ref{sec:hmm-coarse-graining}.  Henceforth, we set  $kT=1$.  At intervals $\tau$, we observe the state of the system and record which well the particle is in.  For now, we assume this measurement is never wrong.

To extract work from a heat bath, we implement the following protocol:  At $t=0$, the potential is symmetric, with no energy-level difference between left and right wells.  We then observe the particle.  If we determine it to be in the right well, then, with no significant time delay, we quickly raise the left well to an energy $E$ (and vice versa if in the left well).  Raising the left well costs no work if we change the potential only where the particle is not present.  From Sekimoto's formulation of stochastic energetics, the work done by an instantaneous change of potential is just $\Delta U$, the change of potential evaluated \textit{at the position of the particle} \cite{sekimoto10}.

We then wait a time $\tau$, keeping fixed the energy $E$ of the left well.  At some time, the particle may spontaneously hop to the left well, because of thermal fluctuations.  At time $\tau$, the left well is quickly lowered back to $E=0$.  If the particle happens to be in the left well, we extract an energy $E$ from the heat bath.  If not, no energy is extracted.  Summarizing, the protocol is to measure the state; then raise the appropriate well by $E$ and wait $\tau$; then lower the well back to 0.

Over many trials, the average extracted work $\langle W \rangle$ is given by $E \, p_\tau$, where $ p_\tau$ is the probability for the particle to be in the left well at time $\tau$.  But $p_\tau$ also depends on $E$.  To evaluate the relation, we consider the \textit{continuous time} dynamics of the state of the system, allowing hops between states at arbitrary times $t$ but still considering the hops themselves to be instantaneous.   The discrete-time master equation $\mathbf{p}_{k+1} = \mathbf{A} \, \mathbf{p}_k$ then becomes $\dot{\mathbf{p}} = \mathcal{A} \, \mathbf{p}$, where the matrix $\mathcal{A}$ has columns that sum to zero, to keep $\mathbf{p}$ normalized at all times.  Normalization implies that a two-state system has but one independent evolution equation, $p(t)$, which obeys
\begin{equation}
	\dot{p} = -\omega_- p + \omega_+ (1-p) \,,
\label{eq:hmm-master-eq}
\end{equation}
where $\omega_-$ is the transition rate out from the left well and $\omega_+$ is the transition rate into the left well.  In equilibrium, detailed balance requires that $\omega_+ / \omega_- = e^{-E}$.  Scaling time so that $\omega_- = 1$ then gives
\begin{equation}
	\dot{p} = - p + e^{-E} (1-p) \,.
\label{eq:hmm-master-eq1}
\end{equation}

Setting $\dot{p}=0$ gives the steady-state solution $p_\infty = 1/ (e^E+1)$.  Notice that $E=0$ implies $p_\infty = \tfrac{1}{2}$, as expected for a symmetric double-well potential, and that $E \to \infty$ implies that the particle is always in the right well ($p_\infty \to 0$).  For finite times, we solve \eref{eq:hmm-master-eq1} with $p_0=0$.  The solution, $p_\tau = p_\infty [ 1 -  e^{-(1+\omega) \tau}]$, implies that
\begin{equation}
	\langle W \rangle = \frac{E}{e^E+1} \, \left[ 1 -  e^{-(1+e^{-E}) \tau} \right] \,.
\label{eq:avg-energy}
\end{equation}
Note that we choose signs so that $\langle W \rangle > 0$ corresponds to work extraction.

Intuitively, for a given cycle time $\tau$, an optimal energy $E^*$ maximizes the average work:  if $E$ is too small, you will extract work in many cycles, but the amount each time will be small.  If $E$ is too large, you will extract more work,  but only very rarely, since the relative probability of being on the left side is $\lesssim e^{-E}$.  For the quasistatic limit $\tau \gg 1$, $\langle W \rangle \approx E/(e^E+1)$, whose maximum $\langle W \rangle^* \approx 0.28$ for $E^* \approx 1.28$.

The second law of thermodynamics implies that $\langle W \rangle \le \Delta F$, where the free energy difference $\Delta F$  is just the difference in entropy $\Delta S$, since the internal energy difference is zero for a cyclic process where the energies of both states are identical at beginning and end.  The maximum entropy difference is $\ln 2 \approx 0.69$, which is considerably larger than the $\approx 0.28$ found in the quasistatic limit of our protocol.

To achieve the $\ln 2$ upper bound for extracted work per cycle, we need to allow $E(t)$ to vary continuously in the interval $0 < t < \tau$ (and to have jump discontinuities at the beginning and end of the interval).  Such continuous-time protocols have been considered previously and lead to  protocols that extract $\ln 2$ of work in the quasistatic limit  \cite{esposito10,diana13,bauer14}.  Nonetheless, we prefer our constant-$E$ protocol:
\begin{itemize}
\item The mathematics is simpler.  The continuous version uses calculus of variations.  The discrete one requires only ordinary calculus.

\item If implemented experimentally, the protocols would almost certainly be carried out digitally, with an output that is fixed between updates.

\item When the goal is to optimize power extraction from the heat bath (rather than work per cycle), the constant-$E$ and continuous protocols give identical results.  
\end{itemize}

To explore this last point, we rewrite \eref{eq:avg-energy} for average power, $\mathcal{P} \equiv \langle W \rangle / \tau$.  Assuming, as a more careful analysis confirms, that maximum average power extraction occurs when $\tau \ll 1$, we have
\begin{equation}
	\mathcal{P} \to \left( \frac{1}{\tau} \right) \, \frac{E}{e^E+1} \, 
		\left[\left(1 + e^{-E} \right) \tau \right] = E \, e^{-E} \,,
\label{eq:avg-power}
\end{equation}
which has a maximum $\mathcal{P}^* = 1/e \approx 0.37$ for $E^*=1$.  The same result is found for the continuous protocol \cite{bauer14}.  Since maximum energy extraction requires quasistatic, infinitely slow manipulations, the power at maximum energy tends to zero.  Maximizing power extraction is arguably more interesting experimentally.

\subsection{Hidden states}

So far, we have assumed noise-free observations.  If the observations are noisy, we have to infer  the probability $p(0) \equiv p_0$ that the particle is in the left well.  Assuming that the particle is likely in the right well ($0 < p_0 < \tfrac{1}{2}$), then we should raise the left well.  After a time $\tau$ has elapsed, \eref{eq:hmm-master-eq1} implies that
\begin{equation}
	p_\tau = p_\infty - (p_\infty-p_0) e^{-(1+\omega) \tau} 
	= \frac{\omega(1-\varepsilon)}{1+\omega}  - \varepsilon p_0 \,,
\label{eq:master-eq-soln}
\end{equation}
with $\omega = e^{-E}$ and $\varepsilon \equiv e^{-(1+\omega) \tau}$.  This expression is linear in $p_0$, as the master equation \eref{eq:hmm-master-eq1} is linear.  The discrete-time master equation for time step $\tau$ then is
\begin{equation}
	 \underbrace{\begin{pmatrix} \alpha & \beta \\ 
	 	1-\alpha & 1-\beta \end{pmatrix}}_{\mathbf{A}_{\rm L}} \,
	 \begin{pmatrix} p_0 \\ 1-p_0 \end{pmatrix} =
	 \begin{pmatrix} \alpha p_0 + \beta(1-p_0)  \\
	 	(1-\alpha) p_0 + (1-\beta) \, (1-p_0) \end{pmatrix} 
	= \begin{pmatrix} \beta + (\alpha-\beta) p_0 \\ (1-\beta) - (\alpha-\beta) p_0 \end{pmatrix} \,.
\end{equation}
Matching terms with $\eref{eq:master-eq-soln}$ gives $\beta = \tfrac{\omega(1-\varepsilon)}{1+\omega}$ and $\alpha = \tfrac{\omega + \varepsilon}{1+\omega}$.  The complements are $1-\beta = \tfrac{1+\omega \varepsilon}{1+\omega}$ and $1-\alpha = \tfrac{1-\varepsilon}{1+\omega}$.  Thus, when the left well is raised, the transition matrix $\mathbf{A}_{\rm L}(E,\tau)$ is 
\begin{equation}
	\mathbf{A}_{\rm L} = \frac{1}{1+\omega} 
		\begin{pmatrix} \omega + \varepsilon & \omega (1-\varepsilon) \\
			1-\varepsilon & 1+\omega \varepsilon \end{pmatrix} \,.
\label{eq:hmm:Maxwell-demon-transition-matrix}
\end{equation}
Notice that the columns of $\mathbf{A}_{\rm L}$ sum to one, as they must and that the Markov transition matrix is no longer symmetric, as expected since we raise one of the wells.  The novel aspect for us is that the transition matrix $\mathbf{A}_{\rm L}$ now depends on the energy level $E$, which can be set at each time step.  

When the right well is raised, matrix elements are switched, with left $\leftrightarrow$ right.  This amounts to swapping ``across the diagonal" of the matrix.  Thus,
\begin{equation}
	\mathbf{A}_{\rm R} = \frac{1}{1+\omega} 
		\begin{pmatrix}1+\omega \varepsilon  &1-\varepsilon \\
			 \omega (1-\varepsilon) & \omega + \varepsilon \end{pmatrix} \,,
\end{equation}

\begin{figure}[ht]
\begin{center}
	\includegraphics[width=5.0in]{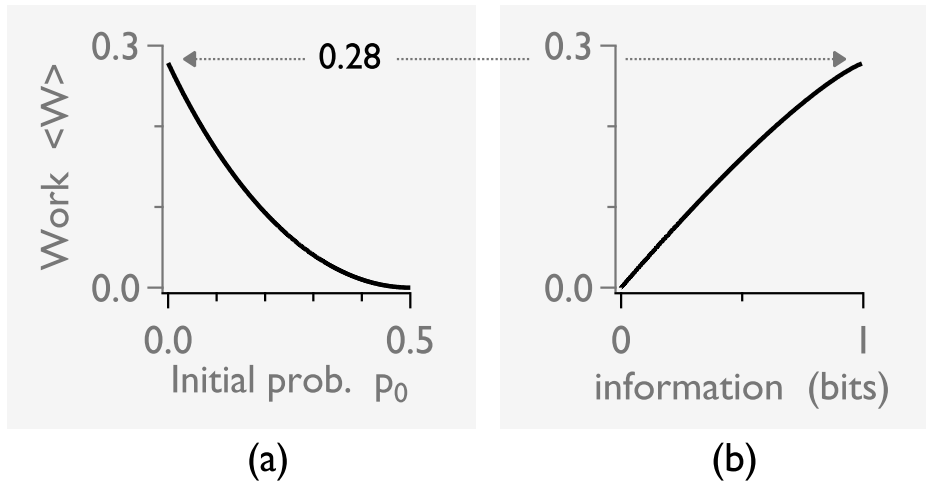}	
	\caption{Maxwell demon extracts work in the quasistatic limit $\tau \gg 1$.  (a)  Average work $\langle W \rangle$ vs. probability to be in the left well at time 0.  (b) Vs. information gain.}
	\label{fig:MaxwellDemonWork}
\end{center}
\end{figure}

The previously analyzed case \eref{eq:avg-energy} for $p_0 = 0$ then represents the best-case scenario:  the particle is definitely on the right, and there is never a penalty for raising the left well.  For $0 < p_0 < \tfrac{1}{2}$, we will occasionally do work in raising the well when the particle is present.  Using \eref{eq:master-eq-soln} and maximizing over $E$, we can quickly calculate the maximum work extraction as a function of $p_0$.  Figure~\ref{fig:MaxwellDemonWork}(a) shows that the maximum average extracted work decreases as the initial state becomes more uncertain.  When $p_0 = \tfrac{1}{2}$, we have no information about the state of the system and cannot extract work from the heat bath, in accordance with the usual version of the second law.  For $p_0 > \tfrac{1}{2}$, we would raise the right well, else we would be erasing information and heating the bath, rather than extracting energy from it.  Figure~\ref{fig:MaxwellDemonWork}(b) shows that the work extracted is nearly a linear function of the change in Shannon entropy between initial and final states.  As in Szil\'ard's analysis, information was used to extract work from the heat bath.  Here, the average slope (converted to nats) gives an efficiency of roughly 41\%.  Less than half the information gained is extracted as work by this particular protocol.

\subsection{Two protocols}

We have not yet specified how to estimate $p_0$ at the beginning of each time interval.  We do so via the observations $y_k$ that are made at the beginning of each control period $\tau$, before the choice of $E$.  The observations have two symbols and are characterized by an observation matrix of the form of \eref{eq:hmm-observation-matrix}, with $b$ the symbol error rate.  We thus return to the formalism discussed in section~\ref{sec:hmm-state-estimation}, where $p_0 \to P(x_k)$, the state of the system at time $k$.  Similarly, $p_\tau \to P(x_{k+1})$.  The only difference is that we modify $\mathbf{A}$ by choosing $E$ and which well to raise at each time step.  Call the choice $\mathbf{A}_k$.

We can incorporate observations in two ways.  One is to use only the observation $y_k$ to estimate $P(x_k)$.  Then Bayes' Theorem implies that $P(x_k | y_k) \propto P(y_k | x_k)$, where the prior $P(x_k) = \tfrac{1}{2}$, since left $\to$ right and right $\to$ left state transitions are equally likely.  Although $P(x_{k+1} | x_k,u_k)$ does not satisfy this condition, the time-averaged sequence of transition matrices does:  since left and right levels are raised at equal frequencies, the overall statistics are symmetric in the absence of other information.  Here, $u_k$ is the control variable, a function of $E$.

The second way is to use the filtering formalism developed in section~\ref{sec:hmm-filtering} to recursively compute $P(x_k | y^k)$.  (Without information about the future, we cannot use smoothing.)  We can say that the second strategy, which depends on past observations, uses memory whereas the first uses no memory.  The procedure is then to
\begin{itemize}
\item Measure $y_k$.
\item Update $P(x_k | y^k)$, based on $\{ \mathbf{A}_k, \mathbf{B} \}$, with the time-dependent transition matrix $\mathbf{A}_k$ given by $P(x_{k+1} | x_k,u_k)$.  The control $u_k$ is a function of $E_k$.
\item Determine $E_{k+1}$ by minimizing $\langle W \rangle (E)$, the average work extracted in a cycle.
\item Apply $u_{k+1}$.
\end{itemize}

\begin{figure}[ht]
\begin{center}
	\includegraphics[width=5.0in]{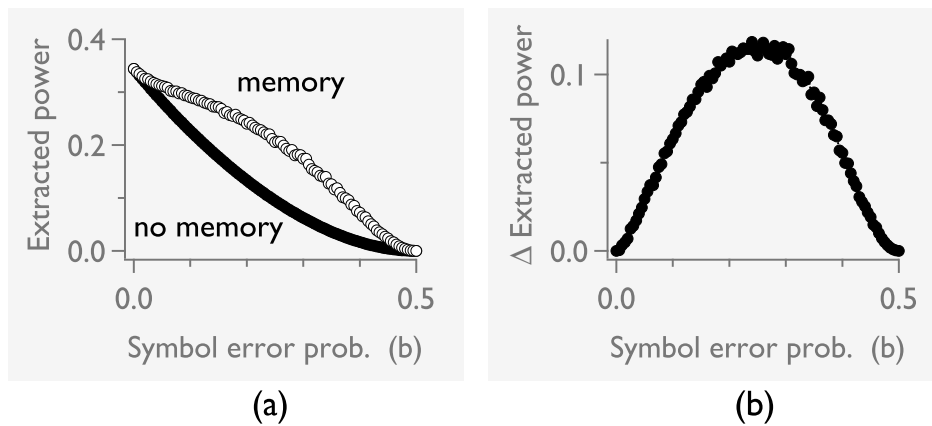}	
	\caption{Maxwell demon extracts power.  (a)  Comparison of power extracted using past and present states $y^k$ to that using only the current state $y_k$. (b)  Difference between the two extracted powers.  Cycle time $\tau = 0.1$.}
	\label{fig:MaxwellDemon}
\end{center}
\end{figure}

Iterated, the above algorithm leads to plots of the average extracted work as a function of the measurement-error probability $b$ (figure~\ref{fig:MaxwellDemon} ).  In (a), the curve labeled \textit{memory}, uses the Bayesian filter to estimate the state of the system.  By ``memory," we mean that the inference about which energy level to alter is based on all the observations $y^k$ up to time $k$.  By contrast, in (b), the ``no memory" curve uses only the current observation, $y_k$.  As before, the extra information from past states is most useful at intermediate values of error rate $b$.  The difference curve, plotted at left below, resembles figure~\ref{fig:FiltervSmoother}, which compared estimator entropies of the smoother and filter state estimates.  The conclusion, again, is that extra information is most useful at intermediate signal-to-noise ratios.  Here, retaining a memory of past observations via the filter allows the Maxwell demon to extract more power from the heat bath.

\subsection{Phase transition in a Maxwell demon}

The continuous-protocol version of the Maxwell demon shows phase transitions in the behavior of the Maxwell demon as the symbol error rate $b$ is varied \cite{bauer14}.  To see that similar phenomena arise in the constant-$E$ protocol discussed in this paper, compare the outcomes of the strategy that uses memory ($y^k$) with one using no memory ($y_k$).  More precisely, we define a ``discord" order parameter $\mathcal{D}$,
\begin{equation}
	\mathcal{D} \equiv 1- \langle y \, \hat{x} \rangle \,, 
\label{eq:discord}
\end{equation}
where $y = \pm 1$ represents the time series of observations and $\hat{x} = \pm 1$ represents the  state estimate, based in this case on the optimal filter.\footnote{
This order parameter has nothing to do with the \textit{quantum discord} order parameter that is used to distinguish between classical and quantum correlations \cite{zurek03}.}
If $y$ and $\hat{x}$ always agree, $\mathcal{D} = 0$.  If $y$ and $\hat{x}$ are uncorrelated, $\mathcal{D} = 1$.  Partial positive correlations imply $0 < \mathcal{D} < 1$.  Put differently, $\mathcal{D}>0$ implies that there is value in having a memory, as the filter estimate $\hat{x}$ can differ from the observation.  When $\mathcal{D}=0$, the filter always agrees with the observation, implying that there is no value in calculating the filter.

\begin{figure}[ht]
\begin{center}
	\includegraphics[width=6.0in]{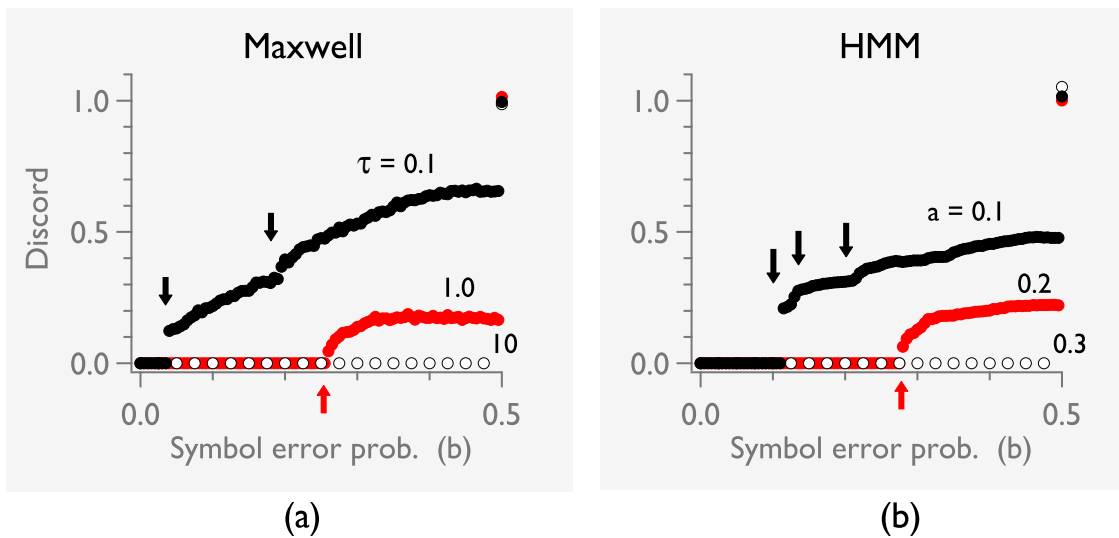}	
	\caption{Phase transition in discord order parameter.  (a)  Maxwell demon, for three different cycle times $\tau$.  Black down-pointing arrows mark jump discontinuities.  Red up-pointing arrow marks a continuous phase transition.  (b) Similar plot for HMM, for three values of transition matrix parameter $a$.}
	\label{fig:MaxwellHMMphaseTrans}
\end{center}
\end{figure}

In figure~\ref{fig:MaxwellHMMphaseTrans}(a), we plot the discord order parameter $\mathcal{D}$ against the symbol error rate $b$ for three different cycle times, $\tau = 0.1$, 1, and 10.  There are many interesting features.  For long cycle times, represented by $\tau = 10$ and hollow markers,  observations match the inferred state---defined here to be the more likely state, as determined by the probabilities from the filter algorithm.  For intermediate cycle times, represented by $\tau=1$ and red markers, there is a continuous bifurcation, or second-order phase transition, indicated by an up-pointing red arrow at $b =b_{\rm c} \approx 0.258$.  (The apparent discontinuity results from the limited resolution of the plot.  At higher resolution, not shown, the bifurcations are clearly continuous.)  For $b < b_{\rm c}$, the filter estimate and observation always agree.  For $b>b_{\rm c}$, they disagree sometimes.  For short cycle times, represented by $\tau = 0.1$ and black markers, we observe two transitions that, upon closer inspection, are both discontinuous, corresponding to first-order phase transitions and marked by down-pointing black arrows.  Finally, at $b=0.5$, the order parameter $\mathcal{D}=1$, since there is no correlation between observation and the internal state (or its estimate).  Interestingly, there is always a jump discontinuity in $\mathcal{D}$ at $b=0.5$.

\section{Phase transitions in state estimation}
\label{sec:hmm-phase-trans}

The phase transition observed in the Maxwell-demon model given in the previous section can also be seen in hidden Markov models that have nothing to do with thermodynamics.  Figure~\ref{fig:MaxwellHMMphaseTrans}(b) shows the discord order parameter $\mathcal{D}$ for a two-state, two-symbol HMM with $x,y \in \{-1,+1\}$, for three values of $a$.  As in figure~\ref{fig:MaxwellHMMphaseTrans}(a), there are first-order transitions for small values of $a$, continuous transitions for intermediate values, and no transitions for larger values.  Intuitively, we need long dwell times in states (low values of $a$) so that we have time to average over (filter) the observation noise.  If so, we may be confident in concluding the true state is different from the observed state.  If the dwell time is short (high value of $a$), the best strategy is to trust the observations.  Note that the values of $a$ correspond roughly to the same regimes as implied by the values of $\tau$; however, we cannot make an exact mapping, since the Markov transition rate in the Maxwell-demon depends on the control $u_k$, which depends on observation errors $b$.

\begin{figure}[ht]
\begin{center}
	\includegraphics[width=6.0in]{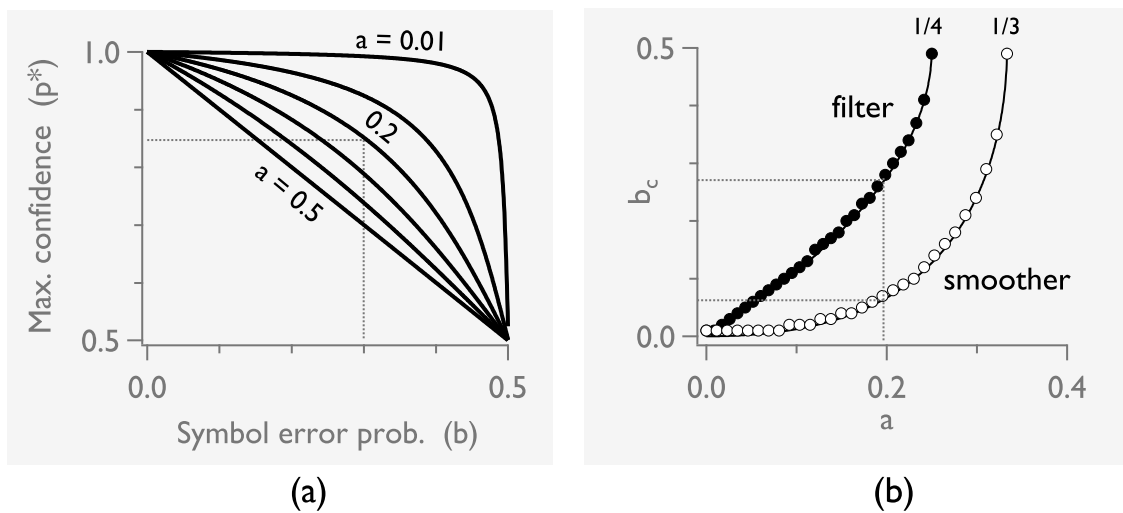}	
	\caption{(a) Maximum confidence level $p^*$ as a function of symbol error probability $b$ for  Markov transition probability $a = $ 0.01, 0.1, 0.2, 0.3, 0.4, 0.5.  Dotted lines show ($a=0.2$, $b=0.3$) case.  (b) Critical value of symbol error probability, $b_{\rm c}$ for filter (solid markers) and smoother (hollow markers), vs.~Markov transition probability $a$.  Simulations as in figure~\ref{fig:MaxwellHMMphaseTrans}(b), with 1000 time units.  For fixed $a$, the parameter $b$ is incremented by 0.01 from 0 until $\mathcal{D}>0.001$, which defines $b_{\rm c}$.  Solid lines are plots of \eref{eq:bcrit-filter} and \eref{eq:bcrit-smoother}.  No parameters have been fit.}
	\label{fig:hmmInftrans}
\end{center}
\end{figure}

As with the Maxwell-demon example, for given $a$ there is a critical value of $b$, denoted $b_{\rm c}$.  To calculate $b_{\rm c}$, we note that there is an upper limit to the confidence one can have in a given state estimate.  As we can see in figure~\ref{fig:hmm-hmmInfFilter}, this limit is achieved after a long string of identical observations, say $y^k=1$, that is $\{y_1=1, y_2=1, \ldots, y_k=1\}$.  See the string of eight $+1$ states in figure~\ref{fig:hmm-hmmInfFilter} as an example.  More formally, we consider $P(x_k=1|y^k=1)$.  For $k \gg 1$, the maximum value of the state probability approaches a fixed point $p^*$ at long times.  The intuition is that even with a long string of $+1$ observations, you cannot be sure that there has not just been a transition and an accompanying observation error.   We derive $p^*(a,b)$ in \ref{sec:appA} and plot the results in figure~\ref{fig:hmmInftrans}(a).

Let us denote $\hat{x}^{({\rm f})}_k \equiv \argmax_{x_k} P(x_k | y^k)$, the filter estimate of $x_k$.  To find conditions where $\hat{x}^{({\rm f})}$ disagrees with $y$, we construct the extreme situation where a long string $y^k=1$ gives the greatest possible confidence that $x_k=1$.  Then let $y_{k+1}=-1$.  The discordant observation must lower the confidence in $x_{k+1}$ to below $\tfrac{1}{2}$ in order for the filter estimate and observation to disagree.  Thus, the condition defining $b_{\rm c}$ is 
\begin{equation}
	P(x_{k+1}=1 | y_{k+1}=-1, y^k=1) = \frac{1}{2} \,.
\label{eq:bcrit-condition}
\end{equation}
Writing this condition out explicitly gives, after a calculation detailed in \ref{sec:appA},
\begin{equation}
	b_{\rm c}^{\rm filter} = \frac{1}{2} \left(1 - \sqrt{1-4a} \right) \,.
\label{eq:bcrit-filter}
\end{equation}
A similar calculation for the smoother, again detailed in \ref{sec:appA}, leads to
\begin{equation}
	b_{\rm c}^{\rm smoother} = \frac{1}{2} \left(1 - \frac{\sqrt{(1+a)(1-3a)}}{1-a} \right) \,.
\label{eq:bcrit-smoother}
\end{equation}

Figure~\ref{fig:hmmInftrans}(b) shows that the thresholds of simulated data agree with \eref{eq:bcrit-filter} and \eref{eq:bcrit-smoother}.  Both filter and smoother estimates imply that there is a maximum value of $a$, call it $a_{\rm c}$, above which $\mathcal{D}=0$ for all $b$.  For the filter $a_{\rm c} = \tfrac{1}{4}$, while for the smoother, $a_{\rm c} = \tfrac{1}{3}$.  The higher value of $a_{\rm c}$ reflects the greater value of smoother vs. filter inferences.

\subsection{Mapping to Ising models}

Although we have explained some features of figure~\ref{fig:MaxwellHMMphaseTrans}, there is clearly more to understand.  For example, there are both continuous and discontinuous transitions, as well as evidence for multiple transitions at fixed $a$.  To begin to understand the reason for multiple phase transitions, we note that the two-state, two-symbol HMM can be mapped onto an Ising model \cite{zuk05,allahverdyan09}.  Let us change variables:
\begin{eqnarray}
	P(x_{k+1} | x_k) &=& \frac{e^{J x_{k+1} x_k}}{2 \cosh J} \,, \qquad 
		J = \frac{1}{2} \ln \left( \frac{1-a}{a} \right) \nonumber \\[3pt]
	P(y_k | x_k) &=& \frac{e^{h \, y_k x_k}}{2 \cosh h} \,, \qquad 
		h = \frac{1}{2} \ln \left( \frac{1-b}{b} \right) \,.
\label{eq:hmm-markov2ising}
\end{eqnarray}
We use these definitions to formulate a ``Hamiltonian" $H = -\ln P(x^N,y^N)$ via
\begin{equation}
	H = -J \sum_{k=1}^N x_k \, x_{k+1} - h \sum_{k=1}^N y_k x_k \,,
\label{eq:hmm-markov2ising-hamiltonian}
\end{equation}
where we have dropped constant terms that are independent of $x_k$ and $y_k$.  For $a < \tfrac{1}{2}$, the interaction term $J>0$ is \textit{ferromagnetic}:  neighboring ``spins" tend to align.  The term $h$ corresponds to an external field coupling constant.  The field $hy_k$ is of constant strength and, for $b < \tfrac{1}{2}$, has a sign is equal to the observation $y_k$.  The picture is that a local,  \textit{quenched} field of strength $hy_k$ tries to align its local spin along the direction defined by $y_k$.  Notice that $h=0$ for $b=\tfrac{1}{2}$:  spins are independent of $y_k$: observations and states decouple.   A further change of variables (gauge transformation), $z_k = y_k x_k$ and $\tau_k = y_k y_{k+1}$, gives 
\begin{equation}
	H(\tau,z) = -J \sum_k \tau_k \, z_k z_{k+1} - h \sum_k z_k \,, 
\label{eq:hmm-markov2ising-hamiltonian1}
\end{equation}
which is a random-bond Ising model in a uniform external field $h$ \cite{allahverdyan14}.

Starting in the late 1970s, both random-bond and random-field one-dimensional Ising chains were extensively studied as models of frustration in disordered systems such as spin glasses.  In particular, Derrida et al. showed that the ground state at zero temperature has a countable infinity of transitions at $h = 2J/m$ for $m = 1, 2, \ldots, \infty$ \cite{derrida78}.  Their transfer-matrix formalism is equivalent to the factorization of the partition function $Z = \prod_k Z_k$ given in \eref{eq:hmm-total-likelihood}.

The lowest-order transition, $h=2J$, corresponds to a case where the external field at a site forces the local spin to align, because we are at zero temperature.  In terms of the original HMM problem, the ground state corresponds to the most likely (Viterbi) path discussed briefly in section~\ref{sec:hmm-state-estimation} \cite{allahverdyan09}.  While the Viterbi path differs from the filter estimate considered here, there may be a similar explanation for the multiple transitions apparent in figure~\ref{fig:MaxwellHMMphaseTrans}.

\section{Discussion}
\label{sec:discussion}

The formalism of hidden Markov models, or HMMs, can both simplify and clarify the discussion of stochastic thermodynamics of feedback using noisy measurements.  Expressed in terms of the control-theory notation developed here, state estimation based on HMM formalism is an effective  way to incorporate the effects of noisy measurements.   As an application, we simplified a previous analysis of a Maxwell demon that uses observations to rectify thermal fluctuations.  We saw that a surprising phase transition in the ``discord" between observation and inferred state is also present in simple HMM models.  At least in this case, the primary source of complexity seems to lie in the process of state estimation, rather than some feature of the thermodynamics.

Our study of phase transitions in the discord parameter follows the methods of Bauer et al. \cite{bauer14}; however, the mathematics is considerably more complicated in that case.  We note that while Bauer et al. do observe a series of transitions in their numerics, they have not seen  evidence for jump discontinuities (private communication).  Perhaps the differences are also associated with the continuous protocol for varying $E$.  More investigation is warranted. 

Beyond simplifying specific calculations, the use of HMMs leads to other insights.  For example, in figure \ref{fig:MaxwellDemon}, we saw that using a memory improves the performance of a Maxwell demon that extracts power from a heat bath.  The greatest improvement was for intermediate values of the noise parameter $b$.  Sivak and Thomson, studying a simple model of biological sensing, reached a similar conclusion \cite{sivak14}.

The results presented here suggest a somewhat broader view.  Figure~\ref{fig:FiltervSmoother} shows a similar result, where the smoother estimate outperforms the filter estimate.  Here, performance is measured by the Shannon entropy of the estimated probability distribution.  Again, we see that the best performance, relative to without memory, is at intermediate noise levels.  Indeed, a variety of similar results can be obtained from many analogous quantities.  For example, filter estimates based on continuous measurements with Gaussian noise also exceed those based on discrete observation measurements, with, again, a maximum at intermediate values of observation noise.

The common feature in all these different examples is that we compute some measure of performance---work extraction, Shannon entropy, etc.---as a function of added information.  This added information can be previous observations (``memory"), offline observations, extra measurement precision, multiple measurements, and so on.  In all cases, the greatest improvement is always at intermediate noise levels or, more precisely, at intermediate levels of signal-to-noise ratio.  Intuitively, the observation makes sense:  if information is perfect (zero noise), then more is superfluous.  If information is worthless (zero signal), then more is again not better.  But in intermediate cases, extra information adds value.  Thus,
\begin{center}
	\fbox{Extra information is most useful at moderate signal-to-noise ratios.}
\end{center}
It would be interesting to try to formalize these ideas further by defining a kind of ``information susceptibility" in terms of a derivative of power extraction, etc. with respect to added information.  In this context, it is worth noting the study by Rivoire and Leibler, who show that the value of information can be quantified by different information theoretic quantities, such as directed and mutual information, when the analysis is causal or acausal \cite{rivoire11}.

Finally, we note that while we have been careful to discuss the smoother as an offline analysis tool whereby data is analyzed after the fact, there are more interesting possibilities.  As stochastic thermodynamics is generalized to accommodate information flows, we should also consider the equivalent to open systems.  For quantities such as energy, we are used to the idea that a subsystem need not conserve energy and that we must account for both energy dissipation and energy pumping.  Analogously, for information, we should consider both dissipation and the consequences of added information.  Because such information comes from ``outside" the system under direct study, causality need not be respected.  For example, consider the problem of controlling the temperature of a house.  A causal control system will simply respond to temperature perturbations after they occur.  If it gets cold, the heater turns on.  On the other hand, we know in advance that at night it gets cold, and we know, with effectively absolute certainty, the time the sun will set.  Thus, we can anticipate the arrival of a cold perturbation and start to compensate for its effects \textit{before} they occur.  The resulting performance gain will be precisely analogous to the results shown in figure~\ref{fig:hmm-hmmInfSmoother}, where we compare filter and smoother estimates.  (The quality of state estimates limits the quality of control.)  

The analysis of noisy discrete dynamics of HMMs is perhaps the simplest non-trivial setting where these ideas may be explored.  More generally, outside influences will appear as additional inputs to a state node in a graphical representation.  In this context, the Bayesian treatment of causality due to Pearl shows how to generalize inferences such as filtering and smoothing to \textit{Bayesian networks}, which have a richer graphical structure than the chain-like Markov and HMMs sketched in figures~\ref{fig:graphicalMarkov}, \ref{fig:hmmFig}, and \ref{fig:graphicalPOMDP} \cite{pearl09,jensen07}.  Such techniques have been used in stochastic thermodynamics to study information thermodynamics on networks \cite{ito13} and would seem to be the right approach to studying systems that are ``causally open."

In conclusion, we have introduced some of the properties of hidden Markov models that make them useful for simplifying the analysis of stochastic thermodynamics in the presence of feedback and noisy measurements, and we have seen how they suggest interesting areas for future research.

\ack

This work was supported by NSERC (Canada).  I thank David Sivak for his helpful comments and review of the manuscript.

\appendix

\section{Calculation of phase transition critical line}
\label{sec:appA}

In the $a$-$b$ parameter plane, the critical line $b_c(a)$ defines the border between the $\mathcal{D}=0$ and $\mathcal{D}>0$ phases.  Informally, the line separates a region where there is no benefit to using the filter estimate from one where there is.  We can use both filter and smoother state estimates to calculate $\mathcal{D}$, giving two different critical lines.

\subsection{Filter case}

For the filter case, we first calculate the maximum confidence $p^*$.  From \eref{eq:hmm-Bayesian-filter},
\begin{equation}
	\underbrace{P(x_k=1|y^k=1)}_{p^*} = \frac{1}{Z_k} \, \underbrace{P(y_k=1|x_k = 1)}_{1-b} 
		\sum_{x_{k-1}} P(x_k=1|x_{k-1}) P(x_{k-1} | y^{k-1}=1) \,.
\label{eq:bcrit-condition00}
\end{equation}
Substituting for the matrix elements in \eref{eq:bcrit-condition00}, evaluating the normalization constant \eref{eq:hmm-partition-function}, and imposing the fixed point gives a quadratic equation for $p^*$:
\begin{align}
	p^* &= \frac{(1-b) \left[ (1-a) \, p^* + a (1-p^*) \right]}
		{(1-b) \left[ (1-a) \, p^* + a (1-p^*) \right] + b \left[(1-a)(1-p^*) + ap^*\right]} \,,
\label{eq:bcrit-condition0}
\end{align}
whose solution is
\begin{align}
	p^* = \frac{1-2b+a (4b-3) + \sqrt{a^2+(1-2a) (1-2b)^2}} {2 (1-2a) (1-2b)} \,.
\label{eq:pstar-soln}
\end{align}

For example, $a=0.2$ and $b=0.3$ gives $p^* \approx 0.852$, which matches the upper bound in figure~\ref{fig:hmm-hmmInfFilter}.  See also figure~\ref{fig:hmmInftrans}(a) in the main text.

In terms of $p^*$, the condition for the threshold $b_{\rm c}$ is given by
\begin{align}
	P(x_{k+1} =1 \, &| \, y_{k+1} =-1, y^k=1) \nonumber \\[3pt]
	&= \frac{P(y_{k+1}=-1| x_{k+1}=1, \cancel{y^k=1}) \, 
		P(x_{k+1}=1|y^k=1)} {P(y_{k+1}=-1| y^k=1)} \nonumber \\[3pt]
	&= \frac{P(y_{k+1}=-1| x_{k+1}=1) \, P(x_{k+1}=1|y^k=1)} 
	{\sum_{x_{k+1}} P(y_{k+1}=-1| x_{k+1}) \, P(x_{k+1} | y^k=1)} \nonumber \\[3pt]
	&=\frac{b[(1-a)p^*+a(1-p^*)]}{b[(1-a)p^*+a(1-p^*)] + (1-b)[ap^* + (1-a)(1-p^*)]} 
		\nonumber \\[3pt] &= \frac{1}{2} \,,
\label{eq:bc-filter0}
\end{align}
Using Mathematica, we reduce \eref{eq:bc-filter0} to 
\begin{align}
	\frac{b \left(1-a+\sqrt{a^2+(1-2a)(1-2b)^2} \right)}
		{(1-2b) \left(1+a-\sqrt{a^2+(1-2a)(1-2b)^2} \right)} = \frac{1}{2} \,.
\label{eq:bc_filter1}
\end{align}
Rearranging and squaring leads to a remarkable simplification,
\begin{align}
	(1-2b)(b^2-b+a) = 0 \,,
\label{eq:bc_filter2}
\end{align}
which has solutions $b=\tfrac{1}{2}$ and $b = \tfrac{1}{2}(1 \pm \sqrt{1-4a})$.  The relevant solution for the phase transition has $b<\tfrac{1}{2}$, which corresponds to the negative root and \eref{eq:bcrit-filter}.

\subsection{Smoother case}

For the smoother, the analogous threshold condition is given by
\begin{align}
	P(x_k=1|y_k=-1,y^{N \backslash k}=1) = \frac{1}{2} \,,
\label{eq:bc_smoother0}
\end{align}
where $y^{N \backslash k} \equiv \{ y_1, y_2, \ldots, y_{k-1},y_{k+1}, \ldots, y_N \} \equiv \{  y^{k-1},y_{k+1}^N \}$, i.e., all the observations except $y_k$.  For the smoother, the future observations are also $+1$.  In words:  if an observation contradicts both past and future, do we trust it?  We write 
\begin{align}
	P(x_k=1|y_k=-1,y^{N \backslash k}=1) = \frac{1}{Z} P(y_k=-1 | x_k=1) \, 
		P(x_k=1 | y^{N \backslash k}=1) \,.
\label{eq:bc_smoother}
\end{align}
We then focus on the second term,
\begin{align}
	P(x_k=1 | y^{N \backslash k}=1)  &= P(x_k=1 | y^{k-1}=1, y_{k+1}^N=1) \nonumber \\[3pt]
	& = \frac{1}{Z} P(y_{k+1}^N=1 | x_k=1,\cancel{y^{k-1}=1}) \, 
		P(x_k=1 | y^{k-1}=1) \nonumber \\[3pt]
	&=  \frac{1}{Z} P(x_k=1 | y_{k+1}^N=1) \, \left( P(y_{k+1}^N=1) / P(x_k=1) \right) \,
		P(x_k=1 | y^{k-1}=1) \nonumber \\[3pt]
	&= \frac{1}{Z} P(x_k=1 | y_{k+1}^N=1) \, P(x_k=1 | y^{k-1}=1) \nonumber \\[3pt]
	&= \frac{1}{Z} P(x_k=1 | y^{k-1}=1)^2 \,,
\label{eq:bc_smoother1}
\end{align}
where we absorb $P(y_{k+1}^N=1)$ and $P(x_k)$ into $Z$ and set $P(x_k | y_{k+1}^N) = P(x_k | y^{k-1})$.  The justification of this last step is that the sole difference in the two conditional probabilities is $P(x_{k+1}|x_k) \to P(x_k | x_{k+1})$.  But these are equal, as Bayes' theorem (or detailed balance) shows:
\begin{align}
	P(x_{k+1}|x_k) = P(x_k | x_{k+1}) P(x_{k+1}) / P(x_k) = P(x_k | x_{k+1}) \,,
\label{eq:bc_smoother3}
\end{align}
where the unconditional probabilities $P(x_k) = P(x_{k+1}) = \tfrac{1}{2}$.

In terms of all these relations, \eref{eq:bc_smoother0} becomes
\begin{align}
	\frac{1}{Z} P(y_k=-1 | x_k=1) \, [P(x_k=1 | y^{k-1})]^2 = \frac{1}{2} \,.
\label{eq:bc_smoother4}
\end{align}
Using our earlier results for the filter, \eref{eq:bc-filter0}, and with $p^*$ given by \eref{eq:pstar-soln},  we have
\begin{align}
	\frac{b \, \frac{(a+p^*-2ap^*)^2}{(a+p^*-2ap^*)^2 + (1-a-p^*+2ap^*)^2}}
		{b \, \frac{(a+p^*-2ap^*)^2}{(a+p^*-2ap^*)^2 + (1-a-p^*+2ap^*)^2} + 
		{(1-b) \, \frac{(1-a-p^*+2ap^*)^2} {(a+p^*-2ap^*)^2 + (1-a-p^*+2ap^*)^2}}}
		 = \frac{1}{2} \,,
\label{eq:bc_smoother5}
\end{align}
Again, an amazing simplification leads to \eref{eq:bcrit-smoother}.  That there are such simple solutions to such complicated equations suggests that a more direct derivation might be found.

\section*{References}


\providecommand{\newblock}{}

\end{document}